\documentclass[conference]{IEEEtran}
\IEEEoverridecommandlockouts
\usepackage{cite}
\usepackage{amsmath,amssymb,amsfonts}
\usepackage{algorithmic}
\usepackage{graphicx}
\usepackage{textcomp}
\usepackage{xcolor}
\usepackage{float}
\usepackage{subfig}
\usepackage{multirow}
\usepackage[hidelinks]{hyperref}

\def\BibTeX{{\rm B\kern-.05em{\sc i\kern-.025em b}\kern-.08em
    T\kern-.1667em\lower.7ex\hbox{E}\kern-.125emX}}
\begin{document}

\title{PATE: Property, Amenities, Traffic and Emotions Coming Together for Real Estate Price Prediction
\thanks{\textsuperscript{*}Corresponding author.\\
\indent This work is supported in part by The National Natural Science Foundation of China (Grant No. 71871006), in part by ACCESS –- AI Chip Center for Emerging Smart Systems, Hong Kong SAR, Hong Kong Research Grant Council (Grant No. 27206321), and National Natural Science Foundation of China (Grant No. 62122004).}
}

\makeatletter
\newcommand{\linebreakand}{%
  \end{@IEEEauthorhalign}
  \hfill\mbox{}\par
  \mbox{}\hfill\begin{@IEEEauthorhalign}
}
\makeatother

\author{\IEEEauthorblockN{Yaping Zhao\textsuperscript{1,2}}
\IEEEauthorblockA{
zhaoyp@connect.hku.hk}
\and
\IEEEauthorblockN{Ramgopal Ravi\textsuperscript{1}}
\IEEEauthorblockA{
raviramgopal@gmail.com}
\and
\IEEEauthorblockN{Shuhui Shi\textsuperscript{1}}
\IEEEauthorblockA{
shishuhui.hit@gmail.com}
\linebreakand
\IEEEauthorblockN{Zhongrui Wang\textsuperscript{1,2}}
\IEEEauthorblockA{
zrwang@eee.hku.hk}
\and
\IEEEauthorblockN{Edmund Y. Lam\textsuperscript{1,2}}
\IEEEauthorblockA{
elam@eee.hku.hk}
\and
\IEEEauthorblockN{Jichang Zhao\textsuperscript{3,*}}
\IEEEauthorblockA{
jichang@buaa.edu.cn}
\linebreakand
\IEEEauthorblockN{\textsuperscript{1}\textit{The University of Hong Kong}}
\and
\IEEEauthorblockN{\textsuperscript{2}\textit{ACCESS–-AI Chip Center for Emerging Smart Systems}}
\and
\IEEEauthorblockN{\textsuperscript{3}\textit{Beihang University}}
}


\maketitle

\begin{abstract}
Real estate prices have a significant impact on individuals, families, businesses, and governments. The general objective of real estate price prediction is to identify and exploit socioeconomic patterns arising from real estate transactions over multiple aspects, ranging from the property itself to other contributing factors. However, price prediction is a challenging multidimensional problem that involves estimating many characteristics beyond the property itself. In this paper, we use multiple sources of data to evaluate the economic contribution of different socioeconomic characteristics such as surrounding amenities, traffic conditions and social emotions. Our experiments were conducted on  $\mathbf{28,550}$ houses in Beijing, China and we rank each characteristic by its importance.
Since the use of multi-source information improves the accuracy of predictions, the aforementioned characteristics can be an invaluable resource to assess the economic and social value of real estate. Code and data are available at: \href{https://github.com/IndigoPurple/PATE}{\textcolor{blue}{https://github.com/IndigoPurple/PATE}}.
\end{abstract}

\begin{IEEEkeywords}
prediction, regression, real estate, traffic, emotion, socioeconomic characteristics, multi-source information, computational social science
\end{IEEEkeywords}

\section{Introduction}
Over the last two decades, models combined with large-scale transaction data have been used by computational social scientists to explain, predict, and even forecast social phenomena. The general aim of real estate price prediction is to identify and exploit socioeconomic patterns emerging in real estate transactions over multiple aspects ranging from the property itself to other factors. We would like to gather variables from different sources and methodological backgrounds to create a model for the purpose of real estate price prediction.


Given the importance of real estate investment to the Chinese economy at 10\% of its Gross Domestic Product (GDP)~\cite{chinesegpa}, an accurate multi-source model that can predict real estate trends will aid the government and private companies to make better informed decisions. Moreover, this will also benefit individuals who are either looking to lease or purchase property. For tenants, rent represents a major monthly expense and individual buyers will look to maximise their long term return on investment. However, real estate valuation is a multifaceted problem that involves estimating many factors of property such as amenities, proximity to public transport and also social emotions. When it comes to data fusion and model selection, these factors make price prediction even more challenging.

Property prices are traditionally estimated based on value of recently sold properties in a given area ~\cite{pagourtzi2003real} and do not consider factors such as surrounding amenities, traffic conditions, proximity to public transport and also social emotions of the location. More recently, the introduction of new sources of data and methods from the computer vision community~\cite{you2017image, poursaeed2018vision}, such as machine learning~\cite{mcgreal1998neural}, have changed traditional methods of property valuation. However, these proprietary systems are heavily dependent on the availability of relevant and updated sales transactions data and fail to take into account the impact of factors such as surrounding amenities, traffic conditions and social emotions on the value of the properties. Although some recent studies incorporate inputs such as infrastructure~\cite{fu2014exploiting}, traffic data~\cite{wardrip2011public}, reviews~\cite{fu2014sparse} and neighborhoods~\cite{fu2014sparse,cortright2009walking,washington2018premium,wang2018urban,hristova2018new,de2018economic}, they rely on limited factors and ignore the complex intricacies associated with a multi-source data model.

In this paper, by exploiting the characteristics of the property together with factors like amenities, traffic and emotions in the surroundings we study how multi-source features influence real estate prices. Specifically, we analyze $28,550$ real estate transactions in Beijing, China~\cite{zhao2022h4m} and provide a multi-modal analysis of the price drivers in play. Our main contributions are as follows:
\begin{itemize}
    \item Our experiments show that the characteristics from different aspects such as amenities, traffic and emotions have an unexpected economic impact on real estate price.
    \item We rank multi-modal features according to their importance to the real estate price, which is quantitatively evaluated by the F-score.
    \item We perform 2 different methods - the linear regression and the XGBoost regression for real estate price prediction. We discover that the use of information from different aspects indeed improves the prediction accuracy of both models.
\end{itemize}

\section{related work}

Estimating the market value of real estate can be viewed as a regression problem with price per square meter being the dependent variable while the independent variables are characteristics that can help correctly determine the price. Therefore the task can be considered a weighted regression of not only property features~\cite{pagourtzi2003real}, historical~\cite{tan2017time} and neighbourhood prices~\cite{hallac2015network} but also pictures~\cite{you2017image,liu2018learning}. For example, You \textit{et al.}~\cite{you2017image} created a Recurrent Neural Network (RNN) using the images of sold houses in the neighborhood. Liu \textit{et al.}~\cite{liu2018learning} combined textual features and external pictures of the sold house to rank and predict the price. Fu \textit{et al.}~\cite{fu2015real} ranked houses through point of interests, their popularity and reviews. Apart from the fact that multimedia data such as pictures requires deep learning to extract features, in fact, existing methods~\cite{madhuri2019house, de2018economic} show that common regression models (\textit{e.g.}, the linear regression and the XGBoost regression) are sufficient to achieve good performance in the house price prediction task.

Conventionally, real estate companies collect historical transaction data, mortgage records and tax assessments before relating these variables to the physical attributes of the property. While this method produces reasonable estimates for buyers and sellers, it completely ignores the importance of other characteristics of a neighbourhood.

Research has been performed to examine the effect of socioeconomic and environmental factors on real estate prices. Cortright \textit{et al.}~\cite{cortright2009walking} found a positive correlation between housing prices and walkability - Homes located in more walk-able neighbourhoods; those with a mix of common daily shopping and social destinations within a short distance command a price premium over otherwise similar homes in less walkable areas. In urban areas, people prefer to rely on public transport rather than cars. Thus the presence of a developed public transportation system and low traffic~\cite{wardrip2011public} in the area can drive up real estate prices. Some researchers also found that intangible qualities of neighbourhoods like a city’s culture~\cite{hristova2018new}, perception~\cite{buonanno2013housing} and design~\cite{poursaeed2018vision} can boost real estate prices. For example, Hristova \textit{et al.}~\cite{hristova2018new} demonstrates that economic capital alone does not explain urban development. By mining posts from Flickr, a photo-sharing site, the authors suggested that the combination of cultural and economic capital is a better indicator of real estate prices and socio-economic conditions. In their article, Buonanno \textit{et al.}~\cite{buonanno2013housing} combined real estate prices and data from a victimization survey to estimate the effect of crime perception on housing prices in Barcelona, Spain. Real estate prices in neighbourhoods that are perceived as less safe than the average for the city are highly discounted. Using vision-based techniques on photos of home exteriors and interiors, Poursaeed
\textit{et al.}~\cite{poursaeed2018vision} developed a model to estimate the luxury level of a property and analysed how this affects the price of real estate. Finally, Boys \textit{et al.}~\cite{smith2017beyond} analysed data from 6 English cities to show how the value of a place is influenced by factors that are not taken into account by conventional economic models. They stated that it is beauty, a sense of local memory and the urban quality of a place that sometimes matter as much or more than connectivity, space and proximity to work. Moreover, their analysis showed that the presence of greenery is not consistently positive as they are normally assumed to be. 

However, each of the aforementioned works only take into consideration a very limited number of factors per time and neglect the role played by other characteristics. And although Nadai \textit{et al.}~\cite{de2018economic} explored many different factors(e.g. security perception, the proximity of greenery) in a neighbourhood, they are quite simple. In this paper we use multiple sources of data to assess the economic contribution of different characteristics such as surrounding amenities, traffic conditions and also social emotions.

\section{method}

\begin{table*}[htbp]
\caption{The name and description of multi-source features we collected and extracted for the real estate data.}
\begin{center}
\begin{tabular}{|c|c|c|l|}
\hline
\#  & Category & Feature & Description\\
\hline
0&  \multirow{8}{*}{Property}  &   Year&     the building year.\\
1&  &    Elvt &      whether there is an elevator in the building.\\
2&   &   RmNum&      the number of bedrooms in the house.\\
3&   &  HllNum&      the number of living and dining rooms in the house.\\
4&   &  KchNum&      the number of kitchens in the house.\\
5&   &  BthNum&      the number of bathrooms in the house.\\
6&    &    Lat&        the latitude of the house.\\
7&    &    Lng&     the longitude of the house.\\
\cline{2-4}
8&  \multirow{10}{*}{Amenity}  &  TspNum&       the number of surrounding transportation infrastructure.\\
9&   &  TspDst&      the average distance of surrounding transportation infrastructure.\\
10&  &  AtrNum&      the number of surrounding tourist attractions.\\
11&  &  AtrDst&      the average distance of surrounding tourist attractions.\\
12&  &  EdcNum&      the number of surrounding education and training institutions.\\
13&  &  EdcDst&     the average distance of education and training institutions.\\
14&  &  HthNum&      the number of surrounding healthcare infrastructure.\\
15&  &  HthDst&        the average distance of surrounding healthcare infrastructure.\\
16&  &  RstNum&      the number of surrounding restaurants.\\
17&  &  RstDst&      the average distance of surrounding restaurants.\\
18&  &  RtlNum&     the number of surrounding retail goods and services.\\
19&  &  RtlDst&      the average distance of surrounding retail goods and services.\\
\cline{2-4}
20&  Traffic &    TrfV&      the average value of daily traffic speeds.\\
\cline{2-4}
21&  \multirow{5}{*}{Emotions} & AgrPct&        the percentage of anger in all emotions.\\
22&  &  DstPct&      the percentage of detestation in all emotions.\\
23&  &  HppPct&      the percentage of happiness in all emotions.\\
24&  &  SadPct&      the percentage of sadness in all emotions.\\
25&  &  FeaPct&        the percentage of fear in all emotions.\\
\cline{2-4}
26& & Price & the price  per square meter of the house in Renminbi (RMB\textsuperscript{a}).\\
\hline
\multicolumn{3}{l}{\textsuperscript{a}RMB is the legal currency of China.}
\end{tabular}
\label{tab:feature_dsc}
\end{center}
\end{table*}

To provide a multi-source data analysis and thus competently predict the real estate price per square meter,  we first perform data collection and pre-processing in Section~\ref{sec:data}. After extracting $27$ features from raw data, we compute feature correlations in Section~\ref{sec:fc}. We then analyze the feature importance to real estate price in Section~\ref{sec:fi}. Finally, we adopt 2 different methods for real estate price prediction in Section~\ref{sec:model} - linear regression and XGBoost regression.

It is worth noting that our objective is to make an attempt to predict real estate prices by combining factors such as property, amenities, traffic and social emotions. Rather than designing and implementing novel prediction methods, we focus on exploring the effectiveness of data fusion and its benefits. As demonstrated in previous work where multi-source data is used for real estate price prediction~\cite{de2018economic}, we adopt the commonly used linear and XGBoost regression as prediction models. Experiment results in Section~\ref{sec:exp} show that these two methods have already achieved feasible performance and sufficiently evaluated the impact of different factors from multi-source data on real estate price.

\subsection{Data}
\label{sec:data}
\subsubsection{Property} 
We collect real estate transaction data from the internet and extract features. Specifically, for each real estate transaction, we extract the basic property features such as the building year and the number of bedrooms. In addition, we obtain the its geographical coordinates from the address of the real estate. In total, we get $28,550$ pieces of real estate data with basic property features.

\subsubsection{Amenities}
Since we know the geographical coordinates for each real estate, we can use Baidu Maps to identify surrounding amenities within a kilometer radius. These are further categorized into $5$ types: transportation, tourist attractions, educational institutions, healthcare facilities and restaurants. We compute the total number and average distance of each type of the surrounding amenities as features of the real estate.

\subsubsection{Traffic}
We also use Baidu Maps to obtain traffic information around every real estate. For each real estate property, we collect its surrounding traffic speed information per $5$ minutes from 6 \textit{a.m} to \textit{12 a.m}. We then compute the average traffic speed value as a feature of the real estate, which intuitively reflects the efficiency of transportation surround the real estate. 

\subsubsection{Emotions}
We collect microblog posts in Beijing and then use the social emotions analysis algorithm from Ref.~\cite{fan2014anger} to analyze the emotions of each microblog post. Each microblog post is classified as one of the following $5$ types: anger, detest (dislike), happiness, sadness, fear. For each real estate, we calculate the percentage of different emotions related to it as emotions features.

In a total, for every real estate transaction, we extract $27$ features of different aspects, \textit{i.e.}, property, amenities, traffic, emotions from the collected multi-source data. Table~\ref{tab:feature_dsc} lists the name and description of those extracted features. Since we aim to predict the prices of real estate, we denote the feature \texttt{Price} in Table~\ref{tab:feature_dsc} as the dependent variable $y$ and the other features as independent variables $x_i (i=0, \dots, 25)$.

\subsection{Feature Correlation}
\label{sec:fc}
Using correlation, we understand the relationship between multiple features. We can use the $r$ value, also called Pearson’s correlation coefficient~\cite{pearson1895correlation}:
\begin{align}
    r_{uv} = \frac{\sum^n_{j=1}(u_j-\overline{u})(v_j-\overline{v})}{\sqrt{\sum^n_{j=1}(u_j-\overline{u})^2} \sqrt{\sum^n_{j=1}(v_j-\overline{v})^2}},
\end{align}
where $n$ is the sample size; $u_j$, $v_j$ are the individual sample points indexed with $j$; $\overline{x}$ is the sample mean obtained by $\overline{x} = \frac{1}{n}\sum^n_{j=1} x_j$; and analogously for $\overline{y}$.

This measures how closely two features are correlated.
The $r$ value is a number between $-1$ and $1$. The closer to $1$, the stronger the positive correlation. The closer to $-1$, the stronger the negative correlation (i.e., the more “opposite” the features are). The closer to $0$, the weaker the correlation. We will discuss the experiments and results in Section~\ref{sec:exp_fc}.

\subsection{Feature Importance}
\label{sec:fi}
Generally, importance provides a score that indicates how useful or valuable each feature is in the construction of the boosted decision trees within the model. The more an attribute is used to make key decisions with decision trees, the higher its relative importance.
This importance is calculated explicitly for each feature, allowing features to be ranked and compared to each other.
We estimate the importance of different features for real estate price prediction problem using the XGBoost library in Python. We will discuss the experiments and results in Section~\ref{sec:exp_fi}.

\begin{figure*}[htbp]
\centering
\vspace{-20mm}
\includegraphics[width=\linewidth]{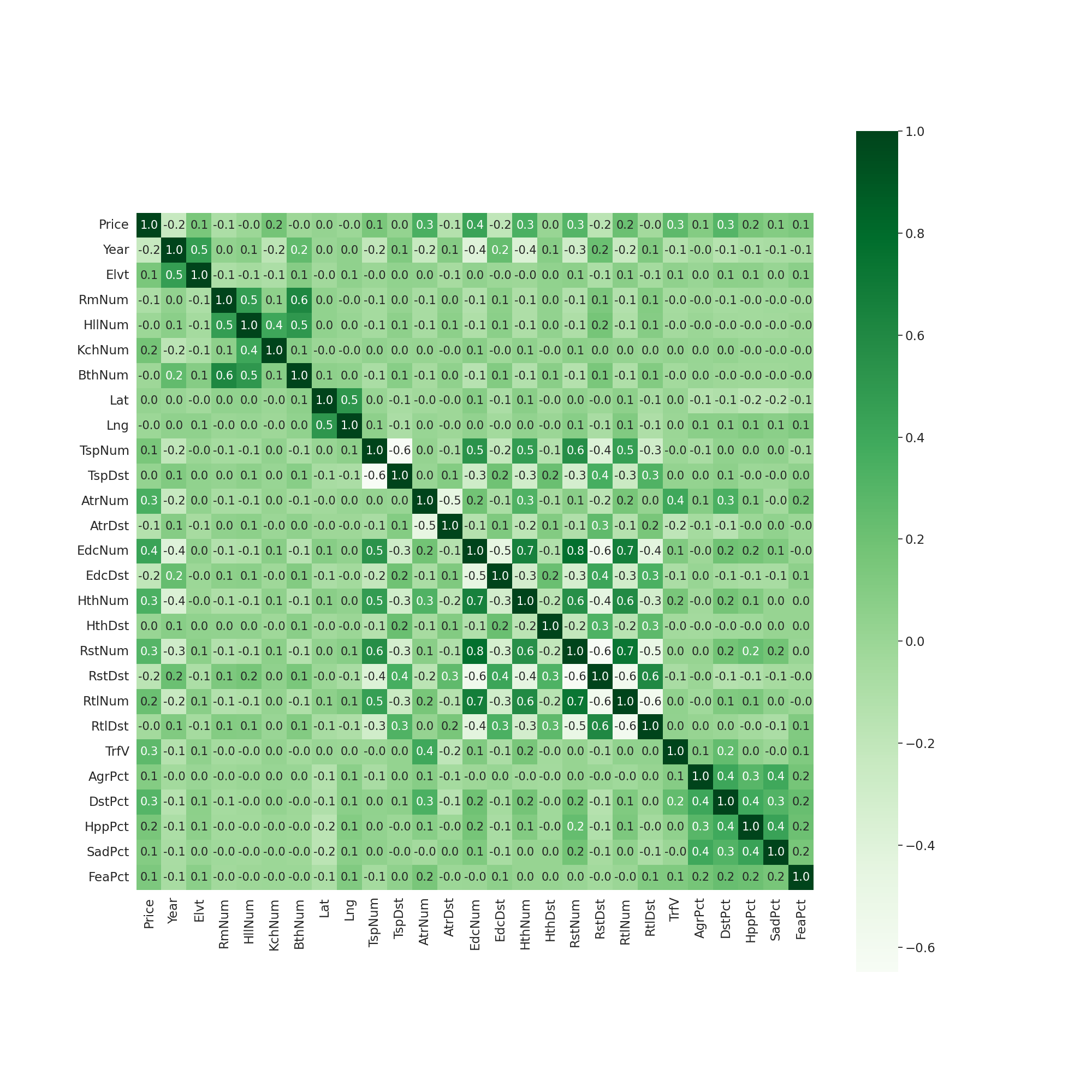}
\vspace{-23mm}
\caption{The relationship between multiple features, measured by the $r$ value, also called Pearson’s correlation coefficient~\cite{pearson1895correlation}. The $r$ value is a number between $-1$ and $1$. The closer to $1$, the stronger the positive correlation. The closer to $-1$, the stronger the negative correlation (i.e., the more “opposite” the features are). The closer to $0$, the weaker the correlation.}
\label{fig:corr}
\end{figure*}

\begin{figure*}[htbp]
\centering
\vspace{-18mm}
\includegraphics[width=\linewidth]{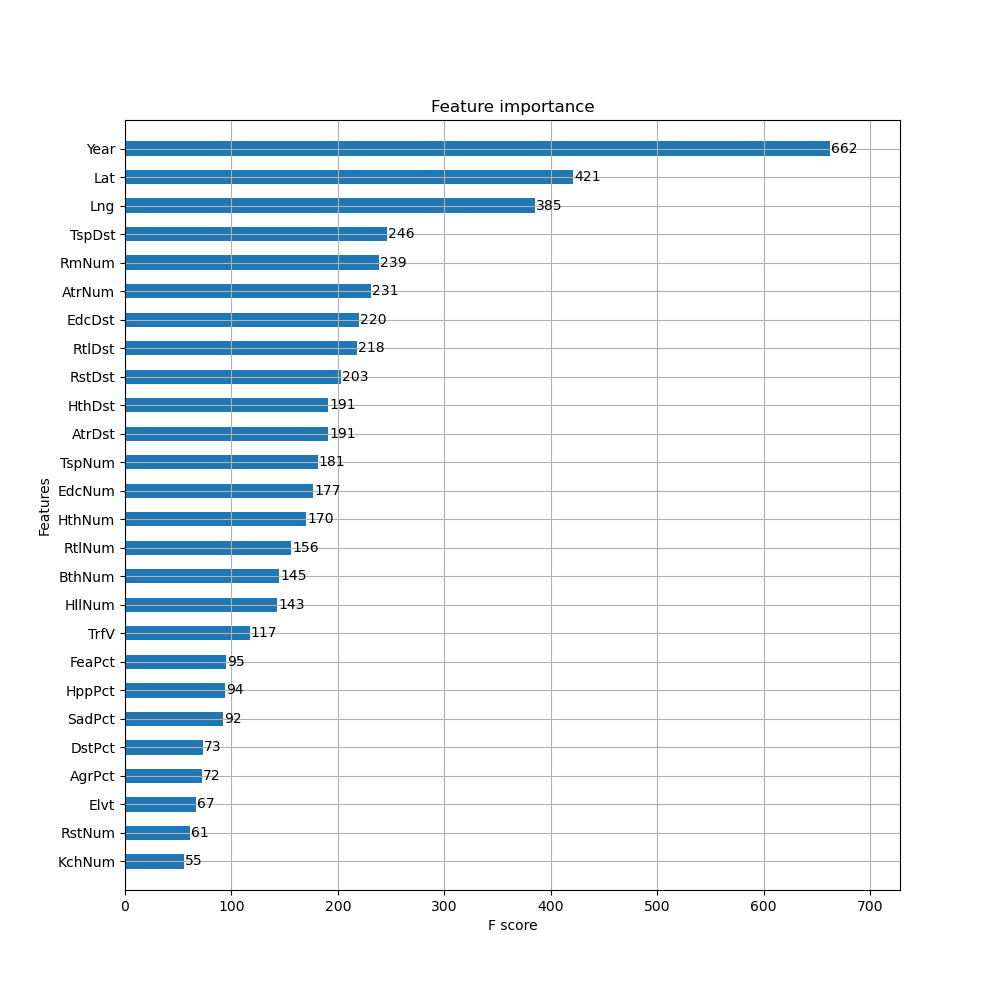}
\vspace{-14mm}
\caption{The importance of different features for the real estate price prediction problem estimated with use of the XGBoost library in Python.}
\label{fig:feature_importance}
\end{figure*}

\begin{figure*}[htbp]
\vspace{-10mm}
    \centering
    \subfloat[]{\includegraphics[width=0.5\linewidth]{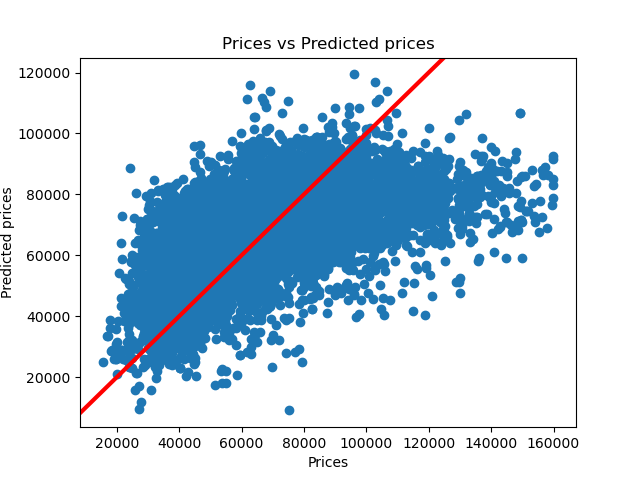}\label{fig:a}}
    \hfill
    \subfloat[]{\includegraphics[width=0.5\linewidth]{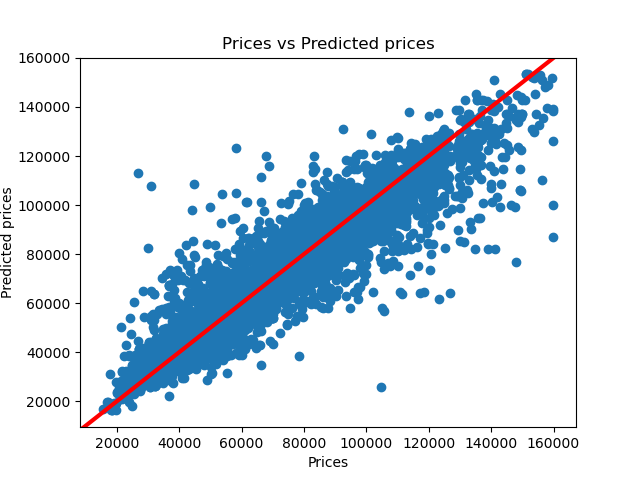}\label{fig:b}}
    \hfill
    \subfloat[]{\includegraphics[width=0.5\linewidth]{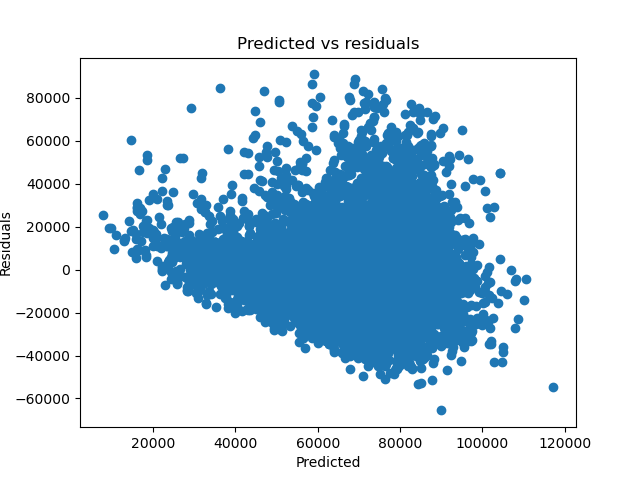}\label{fig:c}}
    \hfill
    \subfloat[]{\includegraphics[width=0.5\linewidth]{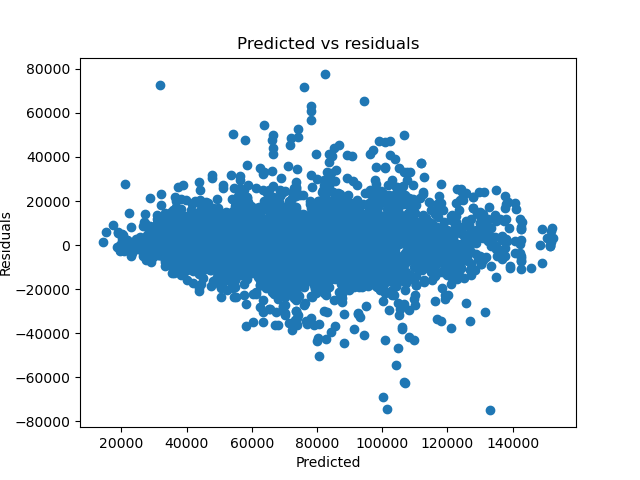}\label{fig:d}}
    \hfill
    \subfloat[]{\includegraphics[width=0.5\linewidth]{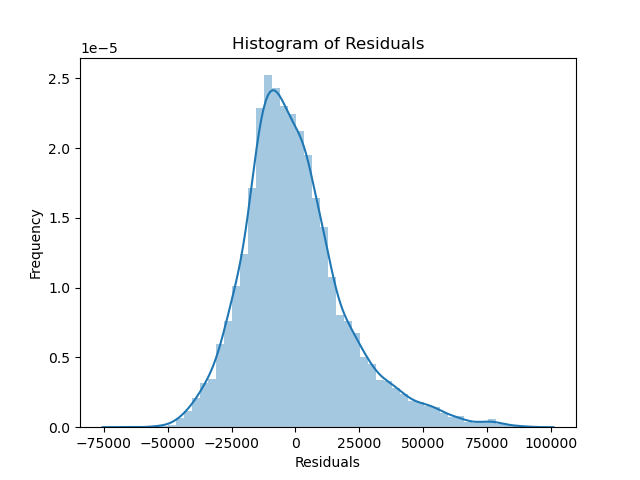}\label{fig:e}}
    \hfill
    \subfloat[]{\includegraphics[width=0.5\linewidth]{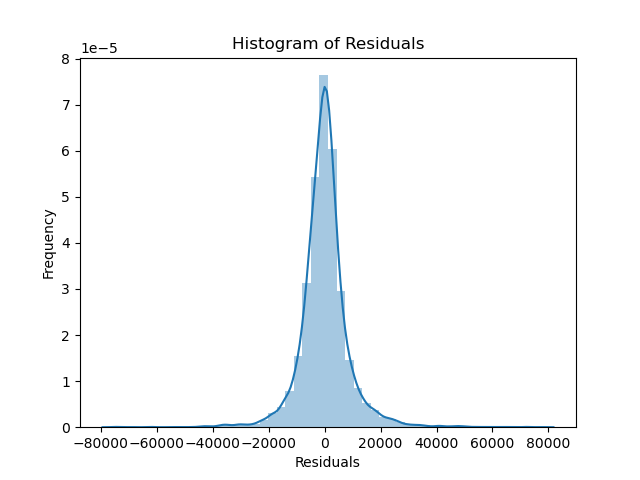}\label{fig:f}}
    \hfill
  \caption{Intuitive comparisons between the linear and XGBoost regression model for real estate price prediction. (a) The differences between actual prices and predicted values of the linear regression model. The red curve is the $x=y$ line. (b) The differences between actual prices and predicted values of the XGBoost regression model. (c) The residuals of actual prices and predicted values of the linear regression model. (d) The residuals of actual prices and predicted values of the XGBoost regression model. (e) The histogram of errors of the linear regression model. (f) The histogram of errors of the XGBoost regression model.}
  \label{fig:lreg}
\end{figure*}

\subsection{Prediction Model}
\label{sec:model}
Using the complex features from multi-source data, we can build models to predict the real estate price. Specifically, we perform two different methods - the linear regression and the XGBoost regression, since prior works~\cite{madhuri2019house, de2018economic} demonstrate that those common regression models are sufficient to yield competent performance in the house price prediction task. 
\subsubsection{Linear Regression}
A linear regression model assumes that the relationship between the dependent variable $y$ and the explanatory variables $x_i (i=0, \dots, 25)$ is linear. Thus the model takes the form:
\begin{align}
    y = \alpha_0 x_0 + \alpha_1 x_1 + \dots + \alpha_{25} x_{25} + \beta,
\end{align}
where $\alpha_0,\dots,\alpha_{25}$ are regression coefficients, and $\beta$ is the intercept term.

\subsubsection{XGBoost Regression}
Due to the simplicity and inaccuracy associated with the linear regression model, We additionally perform an advanced algorithm for real estate price prediction. Extreme Gradient Boosting (XGBoost)~\cite{chen2016xgboost} is an open-source library that provides an efficient and effective implementation of the gradient boosting algorithm. It can be used directly for regression predictive modeling. Therefore, we perform XGBoost regression for advanced prediction modelling.

The loss function of XGBoost is the sum evaluated over all the predictions and a regularization function for all predictors:
\begin{align}
    \mathcal{L}(\theta) = \sum_{j=1}^n l(y_j-\hat{y}_j) + \sum_{t=1}^T \Omega(f_t),
\end{align}
where $n$ is the sample size; $y_j$, $\hat{y}_j$ are the actual and predicted values of the dependent variable; $T$ is the total number of predictors (trees); $f_t$ is the prediction coming from the $t_{th}$ tree. 

We will discuss the experiments and results of the aforementioned two different methods in Section~\ref{sec:exp_pm}.

\subsection{Evaluation}
We use 5 metrics to measure the accuracy of different models: R-squared ($R^2$), adjusted $R^2$, mean absolute error (MAE), mean squared error (MSE), root mean squared error (RMSE). Their formulae are as follows:
\begin{align}
    R^2 &= 1- \frac{\sum_{j=1}^n(y_j-\hat{y}_j)^2}{\sum_{j=1}^n(y_j-\overline{y})^2},\\
    Adjusted R^2 &= 1-[\frac{(1-R^2)\times(n-1)}{n-k-1}],\\
    MAE &= \frac{\sum_{j=1}^n|y_j-\hat{y}_j|}{n},\\
    MSE &= \frac{\sum_{j=1}^n(y_j-\hat{y}_j)^2}{n},\\
    RMSE &= \sqrt{\frac{\sum_{j=1}^n(y_j-\hat{y}_j)^2}{n}},
\end{align}
where $n$ is the sample size; $y_j$, $\hat{y}_j$ are the actual and predicted values of the dependent variable; $\overline{y}$ is the average of all the actual dependent variable values; $k$ is the number of independent variables in the model, excluding the constant.

\begin{table}[htbp]
\caption{The intercept term and regression coefficients of the trained linear regression model.}
\begin{center}
\begin{tabular}{|c|c|c|}
\hline
Intercept & \multicolumn{2}{|c|}{548013.5557669624} \\
\hline 
\#  & Feature & Coefficients\\
\hline
0&       Year&     -225.754\\
1&      Elvt &      8321.2\\
2&      RmNum&      -2341.1\\
3&     HllNum&      -1195.1\\
4&     KchNum&      16563.6\\
5&     BthNum&      4251.97\\
6&        Lat&        10408\\
7&        Lng&     -4548.27\\
8&     TspNum&       28.063\\
9&     TspDst&      25.2256\\
10&    AtrNum&      307.328\\
11&    AtrDst&      3.91726\\
12&    EdcNum&      316.915\\
13&    EdcDst&     -10.4586\\
14&    HthNum&      150.618\\
15&    HthDst&        6.581\\
16&    RstNum&      109.195\\
17&    RstDst&      15.6027\\
18&    RtlNum&     -236.629\\
19&    RtlDst&      4.77269\\
20&      TrfV&      157.894\\
21&    AgrPct&        194.2\\
22&    DstPct&      860.433\\
23&    HppPct&      33.1033\\
24&    SadPct&      35.2334\\
25&    FeaPct&        212.4\\
\hline
\end{tabular}
\label{tab:lreg}
\end{center}
\end{table}

\begin{table*}[htbp]
\caption{Experiments with a variety of settings: 1) using only property features, denoted as \textit{w/ only P}; 2) using property,  traffic, emotions features but without amenity features, denoted as \textit{w/o A}; 3) using property, amenity, emotions features but without traffic features, denoted as \textit{w/o T}; 4) using property, amenity, traffic features but without emotions features, denoted as \textit{w/o S}; 5) using all features including property, amenity, traffic, and emotions, denoted as \textit{w/ PATE}.}
\begin{center}
\begin{tabular}{|c|c|c|c|c|c|c|}
\hline
Data & Method & $R^2$ & Adjusted $R^2$ & MAE & MSE & RMSE \\
\hline
\multirow{10}{*}{Training set} & Linear regression w/ only P & 0.1674 & 0.1671 & 18284 & 551713399 & 23489\\
& Linear regression w/o A & 0.2520 & 0.2515 & 16947 & 495668829 & 22264\\
& Linear regression w/o T & 0.3730 & 0.3722 & 15499 & 415469247 & 20383 \\
& Linear regression w/o E & 0.3636 & 0.3629 & 15582 & 421702012 & 20535\\
& \textbf{Linear regression w/ PATE} & \textbf{0.3797} & \textbf{0.3789} & \textbf{15391} & \textbf{411032381} & \textbf{20274}\\
\cline{2-7}
& XGBoost regression w/ only P & 0.9095 & 0.9095 & 5206 & 59965069 & 7744\\
& XGBoost regression w/o A & 0.9184 & 0.9183 & 4941 & 54069367 & 7353\\
& XGBoost regression w/o T &  0.9331 & 0.9330 & 4416 & 44350499 & 6660\\
& XGBoost regression w/o E & 0.9319 & 0.9318 & 4477 & 45145802 & 6719\\
& \textbf{XGBoost regression w/ PATE} & \textbf{0.9343} & \textbf{0.9342} & \textbf{4387} & \textbf{43549356} & \textbf{6599}\\
\hline 
\multirow{10}{*}{Testing set} & Linear regression w/ only P & 0.1626 & 0.1618 & 17905 & 530648157 & 23036\\
& Linear regression w/o A & 0.2437 & 0.2424 & 16696 & 479250332 & 21892\\
& Linear regression w/o T & 0.3591 & 0.3572 & 15324 & 406118449 & 20152 \\
& Linear regression w/o E & 0.3510 & 0.3494 & 15358 & 411213070 & 20278\\
& \textbf{Linear regression w/ PATE} & \textbf{0.3651} & \textbf{0.3632} & \textbf{15235} & \textbf{402302484} & \textbf{20057} \\
\cline{2-7}
& XGBoost regression w/ only P & 0.8560 & 0.8558 & 6244 & 91267181 & 9553\\
& XGBoost regression w/o A & 0.8646 & 0.8644 & 6080 & 85802314 & 9263\\
& XGBoost regression w/o T & 0.8751 & 0.8747 & 5773 & 79153827 & 8897\\
& XGBoost regression w/o E & 0.8740 & 0.8737 & 5814 & 79830419 & 8935\\
& \textbf{XGBoost regression w/ PATE} & \textbf{0.8770} & \textbf{0.8766} & \textbf{5721} & \textbf{77956264} & \textbf{8829}\\
\hline
\end{tabular}
\label{tab:exp}
\end{center}
\end{table*}

\section{experiment}
\label{sec:exp}

\subsection{Feature Correlation}
\label{sec:exp_fc}
As Figure~\ref{fig:corr} shows, the $r$ value of correlation between the \texttt{Price} and other variables is within the range of $[-0.2, 0.4]$ thus indicating that the linear correlation is weak. Therefore, it is unlikely to use only one or few features to accurately predict the real estate price. Rather, it is necessary and effective to collectively gather multiple features of different aspects from multi-source data for accurate prediction.

Besides, considering the correlation to the real estate price, we additionally observe that features within the same category would show a certain correlation. Specifically, we observe some interesting patterns:
\begin{enumerate}
    \item For the property features, the correlation coefficient on the \texttt{RmNum} and \texttt{BthNum} is $0.6$. In other words, the number of bedrooms in the house and the number of bathrooms have high correlation.
    \item For the amenity features, they show a lot of different relationships. For example, the correlation coefficient on the \texttt{EdcNum} and \texttt{HthNum} is $0.7$, on \texttt{EdcNum} and \texttt{RstNum} it is $0.8$ and finally on \texttt{Rst} and \texttt{RtlNum}, it is $0.7$. Those observations indicate that educational institutions often have healthcare facilities and restaurants nearby. Moreover, restaurants and retail goods and services are together in the same area.
    \item The traffic feature \texttt{TrfV} shows a relatively strong relationship with the \texttt{AtrNum}. This indicates that vehicular traffic is high in areas with high number of tourist attractions.
\end{enumerate}

The above observations suggest several open questions for future work.

\subsection{Feature Importance}
\label{sec:exp_fi}
Figure~\ref{fig:feature_importance} ranks the importance of different features by their $F$-scores~\cite{van2004geometry} in the XGBoost model. We could observe that: 
\begin{enumerate}
    \item The building year of the real estate, denoted as \texttt{Year}, \textit{i.e.}, the age of the house is the most important factor for the price.
    \item The location of the real estate, represented by its latitude \texttt{Lat} and longitude \texttt{Lng}, is the second most important factor for the price.
    \item Among all the amenity-related features, the average distance of surrounding transportation infrastructure \texttt{TspDst} is the most important. This indicates that the transport convenience is an important factor for consumers who are looking to buy real estate.
    \item The average value of daily traffic speeds \texttt{TrfV} ranks 18th of all the $26$ features. This indicates that even though \texttt{TspDst} and \texttt{TrfV} are both factors related to traffic convenience, real estate consumers care more about the proximity to public transport and relatively ignore the degree of surrounding traffic congestion.
    \item As for the structure of the real estate, the number of bedrooms in the house \texttt{RmNum} has the greatest impact on the price, and far exceeds the impact of other types of rooms (\textit{e.g.}, bathroom, living room and kitchen). The number of kitchens \texttt{KchNum} ranks the last among all features, indicating that consumers usually do not care about the number of kitchens in a house. 
    \item Another interesting observation is that the number of tourist attractions \texttt{AtrNum} ranks second in the amenity-related features and the sixth in all features thus indicating that the real estate in areas with more tourist attractions would have higher price.
    \item As for the social emotions around the real estate, we found that fear, happiness and sadness (\textit{i.e.} \texttt{FeaPct}, \texttt{HppPct}, \texttt{SadPct}) have greater impact on the price than detest and anger (\textit{i.e.}, \texttt{DstPct}, \texttt{AgrPct}).
\end{enumerate}

\subsection{Prediction Model}
\label{sec:exp_pm}
In total we have $28,550$ pieces of data with a dependent variable $y$ and $26$ independent features $x_i (i=0, \dots, 25)$ as the dataset. We randomly split the dataset and then use $70\%$ of the data as the training set and the remaining $30\%$ as the testing set.
Then we perform two different methods for real estate prediction and discuss their results as follows.

After training, we obtain a trained linear regression model as Table~\ref{tab:lreg} shows and also a trained XGBoost regression model. For an intuitive comparison of linear and XGBoost regression models, we visualize their results on the testing set in Figure~\ref{fig:lreg} and observe the following:
\begin{enumerate}
    \item Figure~\ref{fig:a}, \ref{fig:b} show the differences between actual prices and predicted values of the linear and XGBoost regression model respectively. While the predicted values of the linear regression model tend to be higher than the actual prices, the data points of the XGBoost model evenly distribute around the $x=y$ line thus demonstrating that the prediction of the XGBoost model is accurate.
    \item Figure~\ref{fig:c}, \ref{fig:d} show the residuals of actual prices and predicted values of the linear and XGBoost regression model. Values of the XGBoost model are distributed equally around zero while the linear regression model has a lot of obvious outliers especially around the predicted value $= 80000$.
    \item Figure~\ref{fig:e}, \ref{fig:f} shows the histogram of errors of the linear and XGBoost regression model. The residuals are normally distributed for both models. However, the variance of the linear regression errors are larger than that of the XGBoost. 
\end{enumerate}

\subsection{Ablation Study}

To evaluate the impacts of different factors from multi-source data on the real estate price, we observe the performance of the linear and XGBoost regression models on the training and testing datasets. 

Specifically, we perform experiments with a variety of settings: 1) using only property features, denoted as \textit{w/ only P}; 2) using property, traffic, emotions features but without amenity features, denoted as \textit{w/o A}; 3) using property, amenity, emotions features but without traffic features, denoted as \textit{w/o T}; 4) using property, amenity, traffic features but without emotions features, denoted as \textit{w/o E}; 5) using all features including property, amenity, traffic, and emotions, denoted as \textit{w/ PATE}.
 
As Table~\ref{tab:exp} shows, compared to the \textit{w/ only P} using only the traditional property features, adding any additional feature (\textit{i.e.}, amenity, traffic, or emotions) could improve the performance. We also observe that: 1) Using all features from different aspects including property, amenity, traffic, and emotions achieves the best performance. 2) Removing the amenity feature leads to the largest performance drop, compared to removing the traffic or emotions. This indicates that the amenity features have a greater impact than the traffic and emotions 3) Removing the traffic feature has a relatively slight impact on the performance. It is reasonable since the traffic feature has only one dimension. However, using the one-dimensional traffic feature still could improve the performance.

\section{discussion}
Taken together, our results clearly show the economic effect of different aspects on real estate price. From our analyses, we may also draw several implications for citizens, local authorities and urban planners, and for real-estate investors:

\begin{enumerate}
    \item \textit{For individuals:} Buying property is a significant investment for people. Therefore in addition to characteristics of the house, they should not only take into consideration other facets such as proximity to amenities and public transport but also intangible aspects such as the perception of a city’s culture and security. A model which helps make an informed decision using the above factors will lead to improved quality of life due to improved ``walkability'' and less traffic and also increase the probability of a significant return on the original investment. 
    \item \textit{For government authorities and urban planners:} For government authorities and urban planners: Understanding the importance of different features in real estate price prediction can help planners and local authorities with urban development. For example, from Figure 2 we can interpret that the proximity to amenities and services such as public transport and schools has a positive influence on real estate prices. Therefore infrastructure projects and developing public spaces will help refine certain areas of the city. Additionally, by studying trends in changes to the built environment can help identify negative impacts of gentrification such as forced displacement, a fostering of discriminatory behavior by people in power, and a focus on spaces that exclude low-income individuals and people of color. Data models such as PATE are computational tools that will help analyse cities.
    \item \textit{For Property developers:}  Rather than just focusing on the characteristics of houses and apartments such as size, number of bedrooms etc, property developers need to expand their view to include place and surroundings. Ideally, urban planners and developers should work in tandem to achieve a high urban vitality. For example, Figure 2 suggests that people would prefer to live in areas with pedestrian spaces and a variety of establishments in its vicinity. In contrast, areas with few amenities and surrounded by large streets that make pedestrian movements difficult have a low vitality.
\end{enumerate}

\section{future work}
In this paper, our goal is to make an initial attempt to fuse multi-source data for real estate price prediction. Rather than designing and implementing novel prediction methods, we adopt the commonly used linear and XGBoost regression as prediction models. This paper opens possibilities of future work. We could attempt to further improve the prediction accuracy and also enhance performance by designing complex variants of existing models. To stimulate further research, the data and code are available publicly at: \href{https://github.com/IndigoPurple/PATE}{\textcolor{blue}{https://github.com/IndigoPurple/PATE}}. While we aim to use our data to enable positive contributions to real estate price prediction, it can also be used to explore its relevance in other applications such as traffic congestion prediction~\cite{zhou2022traffic}.

\section{conclusion}
In this work, we have studied the economic impact of multi-aspect characteristics on real estate price. This analysis was formulated as a multi-modal problem where we predict the price of real estate given conditions from multi-source data. We collected and re-processed a large dataset of real estate properties and computed surrounding amenity, traffic, and emotions features.
We trained and tested the linear and XGBoost regression algorithms and found that the use of information from different aspects increases the prediction accuracy of the model.
Our experiments show that the characteristics from different aspects \textit{i.e.} amenities, traffic, emotions, have an economic impact on real estate price.
We ranked multi-modal features according to their importance to the real estate price.

Based on our observations in Section~\ref{sec:exp_fc}, our work suggests several open questions for future work. We hope that this research and released methods can pave the way for novel studies on the previously neglected link between the cost of houses and many other factors.

\newpage
{
\bibliographystyle{unsrt}
\bibliography{egbib}
}

\end{document}